\documentclass[a4paper]{article}
\usepackage{amsmath}
\usepackage{graphicx}
\usepackage{geometry}
\usepackage{floatrow}
\usepackage{layout}
\usepackage{amssymb} 
\usepackage{multirow}
\usepackage{caption}
\usepackage{authblk}
\usepackage{indentfirst}
\usepackage{url}
\usepackage{hyperref}
\usepackage[
style=ieee
]{biblatex}
\begin{document}
\title{\textbf{\huge{Comparing Classification Models on Kepler Data }}}
\author{\textbf\large{Rohan Saha - rsaha@ualberta.ca}}
\affil{\textbf{University of Alberta}}
\date{November 26, 2019}
\maketitle
\maketitle
\section{Introduction}
\label{sec:Intro}
Even though the original kepler mission ended due to mechanical failures, the kepler satellite continues to collect data. 
Using classification models, we can understand the features exoplanets possess and then use those features to investigate further for any more information on the candidate planet. Based on the classification model, the idea is to find out the probability of the planet under observation being a candidate for an exoplanet or a false positive. If the model predicts that the observation is a candidate for being an exoplanet, then further investigation can be conducted. From the model we can narrow down the features that might explain the difference between a candidate and false positive which ultimately help use to increase the efficiency of any model and fine tune the model and ultimately the process of searching for any future exoplanets.

\section{Dataset Description}
\label{sec:dataset_description}
Space agencies have placed telescopes in the orbit to look for exoplanets. Exoplanets are planets that are not present in our solar system. NASA had launched the telescope in 2009 for finding exoplanets in other star systems, with the goal of finding other habitable planets, which might be explored in the future with scientific advancements. Though the mission ended due to mechanical failures, the satellite still records images on an extended mission.
The data set contains numerous features which are explained by the data dictionary \href{https://exoplanetarchive.ipac.caltech.edu/docs/API_kepcandidate_columns.html}{\textcolor{blue}{here}}. Each sample in the dataset is termed as a Kepler “object of interests” or KOIs. The data set contains 9564 samples and 49 columns as features and one target variable which is categorical. Some columns of interests for the experiment are given below\cite{noauthor_kepler_nodate}:
\begin{itemize}
    \item kepler\_name: [These names] are intended to clearly indicate a class of objects that have been confirmed or validated as planets—a step up from the planet candidate designation. However, this attribute will not be included for classification.
    \item koi\_pdisposition: The disposition Kepler data analysis has towards this exoplanet candidate. One of FALSE POSITIVE, NOT DISPOSITIONED, and CANDIDATE.
    \item koi\_score: A value between 0 and 1 that indicates the confidence in the KOI disposition. For CANDIDATEs, a higher value indicates more confidence in its disposition, while for FALSE POSITIVEs, a higher value indicates less confidence in that disposition.
\end{itemize}
In addition, features like koi\_fpflag\_nt and koi\_fpflag\_ss have high correlation with the target variable, where the former deals with the light curve of the kepler object of interest and the latter deals with the transit-like event and whether the transit like event was significant. From preliminary exploratory data analysis, these two features had high scores in terms of their relation with the target variable. Also, koi\_depth is an important variable for problem and its relation related to koi\_duration and how these two variables affect the candidacy of the astronomical body being an exoplanet.

\section{Problem Description}
\label{sec:Sec3}
For the purposes of this problem, the target variable under consideration is the ‘koi\_disposition’, and with a quick inspection of the data set, it is observed that there are two categorical outputs, “CANDIDATE”, and “FALSE POSITIVE”. Therefore, is a binary classification problem. The former output says that the sample under observation is a candidate for being an exoplanet and the latter output says that the sample under observation is false positive for being considered as a planet. This poses an important question of finding out how similar should be the attributes so as to distinguish between false positives and candidates. This is of paramount importance because if a model predicts a sample as a candidate then the decision of undertaking further investigation should be carried out for the sample to be confirmed as an exoplanet; this approach will filter out observations ultimately saving time. 
Some observations after looking at the data are as follows. \\
\begin{enumerate}
    \item The number of samples are almost equally distributed for each label.\\
		a. There are 4,496 candidate exoplanets\\
		b. There are 5,068 false positives
	\item koi\_depth - Candidate exoplanets tend to have a lower koi\_depth compared to False Positives. However, the distinction is not clear because there are also some exoplanets that have a higher koi\_depth. Therefore, predicting the label solely based on the koi\_depth will be a naive decision. koi\_dpeth is the fraction of stellar flux lost at the minimum of the planetary transit.
\end{enumerate}

\section{Experimental Design}
\label{sec:Sec4}
The experiment under consideration was divided into multiple parts for ease of understanding and clarity. The choice of algorithms and justification is provided in a later subsection. \\

\subsection{Exploratory data analysis}
\begin{figure}[h!]
    \centering
    \includegraphics[width=0.6\linewidth]{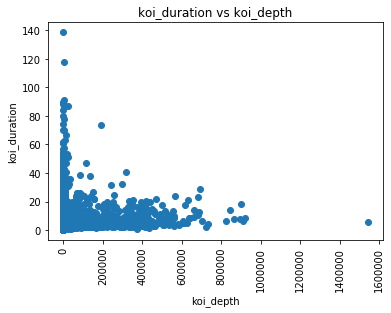}
    \caption{koi\_duration vs koi\_depth}
    \label{fig:eda_koi_depth_koi_duration}
\end{figure}

Before preprocessing, basic exploratory data analysis was carried out to obtain a high level understanding of the data. We first looked at the proportion of the number of samples and found out that the dataset has almost equal number of samples for each class (CANDIDATE and FALSE POSITIVE). And from some preliminary data analysis the most insightful observation was from the two variables \texttt{koi\_duration} and \texttt{koi\_depth}. These two variables are most insightful because they represent the transit attributes of the astronomical body under consideration. \texttt{koi\_duration} represents the time taken from the start to the end when the astronomical body passes in front of the host star\cite{noauthor_nasa_nodate} and \texttt{koi\_depth} represents the fraction of stellar flux lost at minimum during transit. Figure \ref{fig:eda_koi_depth_koi_duration} shows that low \texttt{koi\_depth} is significant because lesser amount of stellar flux lost represents that the astronomical body is closer to the host star and thus has a high possibility of revolving around an orbit. Low \texttt{koi\_duration} signifies that transit time is less and has a higher chance of being within the gravitational influence of the host star. Therefore, low values of these two parameters support candidacy of the astronomical body for an exoplanet. \\
From figure \ref{fig:eda_koi_depth_koi_duration}, it can be observed that candidate exoplanets usually have low koi\_depth and low koi\_duration. On the other hand, false positives either have high koi\_depth and low koi\_duration or low koi\_depth and high koi\_duration.

\newpage
\subsection{Feature Selection}
The first part involved feature selection, where we remove the features which are redundant. Out of the initial 49 feature variables, 18 were removed. This reduce the dimensionality of the data. The features were removed manually after calculating the pearson correlation between the features. After calculating the correlation, the dependent features were removed from the dataset\footnote{The $\chi^2$ method for selecting best features could not be used because the dataset contains negative values, which could have been solved with normalizing the features between 0 and 1. However, it was out of the scope of the experiment.}. It must be noted that the features were selected manually because the number of features is small enough and it also helps us to understand the data at a higher level.\\
The scores for the top features are given in figure 
\begin{figure}[h!]
    \centering
    \includegraphics[width=0.7\linewidth]{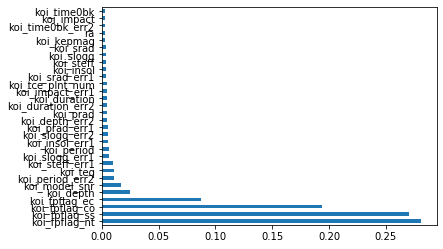}
    \caption{Feature Scores for top 30 features}
    \label{fig:features_scores}
\end{figure}

\subsection{Preprocessing}
\label{sec:sec4.3}
The second part is to preprocess the dataset. From initial examination of the dataset, there were considerable number of samples, which contained null values. One could simply ignore the samples from the dataset while training the model. However, such an approach is valid when the number of samples containing null values are extremely small. This ensures that there is no significant loss of information when the samples are dropped. In the case where there is considerable amount of samples, dropping samples leads to loss of information and thus requires replacing the null values in the samples. Here, we will learn the distribution over the non-null values in the samples in order to obtain the value of the feature containing the null value. This is analogous to learning the distribution over the target variables given the feature values in the dataset. The iterative imputer from the scikit-learn package will be used to accomplish the task\footnote{IterativeImputer is an experimental feature in the scikit-learn package and must be used with caution.\\ \url{https://scikit-learn.org/stable/modules/generated/sklearn.impute.IterativeImputer.html}}. The iterative imputer calculates p(x$\vert$y) where x is the value of the feature which has missing value and y contains the rest of the features which contain values. \\
After the dataset has been preprocesed, the dataset has to be divided into training and testing data. For the purposes of the experiment, the dataset has been divided in an 80-20 split where 80\% of the data will be used for training and 20\% will be used for testing, which gives us 7651 samples for training and 1913 samples for testing. Also, cross-validation\cite{noauthor_cross_validation_nodate} will be used with k=5 and the validation set for the cross-validation will be a part of the training set. The test set will never be used for any form of training as this might bias the results and result in overfitting. Also, no sample from the testing set is included in the training process in any form. This is ensured by implementing cross-validation after the training-test dataset split.

\subsection{Algorithms}
\label{sec:sec4.4}
The algorithms that will be used for modelling the dataset will be 
\begin{enumerate}
    \item Logistic Regression
    \item Decision Tree
    \item Artificial Neural Network
\end{enumerate}

\subsection{Choice of Algorithms}
\label{sec:choice_of_algorithms}
\subsubsection{Logistic Regression}
Logistic Regression is one of most basic models which is used in machine learning for classification. The term logistic refers to the logistic function that helps to normalize the output between the values 0 and 1. This helps to set a threshold (usually 0.5) above which the class for the sample is classified to be one of the classes(usually class one) and below which the sample is classified to be in the other class (usually class zero). In other words, logistic regression tries to maximize the posterior class probability. The logistic function is the sigmoid function which normalizes the output between zero and one. The sigmoid function is given as:
\begin{align}
    \sigma(x) = \frac{1}{1+\exp^{-x}}
\end{align}
In fact, a neural network also uses a sigmoid function in the output layer in case of binary classification which we will examine later. \\

The meta-parameter under consideration for logistic regression was the \textbf{inverse regularization} parameter for which the values were in the set \{0.001, 0.01, 0.1, 1, 10, 100, 1000\}.

\subsubsection{Decision Tree}
The initial choice was to use support vector machine for the problem. However, we found out that the Decision Tree classifier also provides similar performance to that of support vector machine. The initial choice of SVM was motivated by the fact that it is robust to outliers but decision tree is also robust to outliers. For the dataset, SVM took much longer to train than Decision Tree, but performed similarly. Therefore, decision tree was chosen as one of the algorithms for this experiment.
Also, Decision Tree is a simpler model than Support Vector Machine in terms of intuition and understanding.
A high level understanding of Decision Trees can be found out \href{https://medium.com/@simpleparadox/decision-trees-branching-gives-you-wings-part-1-samur-ai-2d943f1affb7}{\textcolor{blue}{here}}, and a more in depth understanding is provided \href{https://medium.com/greyatom/decision-trees-a-simple-way-to-visualize-a-decision-dc506a403aeb}{\textcolor{blue}{here}}. \\
For the Decision Tree algorithm, the meta-parameter under consideration was the depth of the tree\cite{decision_tree}. \\
The values under observation for the tree depth $\in \mathcal \mathbb{Z^+}$ and lie in the set \{1,2,3,...,25\}. It was also observed that higher values of depth did not have any improvement in the performance of the decision tree model.

\subsubsection{Artificial Neural Network}
Neural Networks are a variations of generalized linear models that learn a chain of hidden representations with the help of non-linear activations. For the purpose of the experiment, a two layered neural network was used with 10 hidden units in each hidden layer and with ReLU activation in the hidden layer. The rectified linear unit is used because is is faster in terms of the training process and intuitive. The last layer had a sigmoid activation and the loss function used was binary cross-entropy or the log loss, which is the same loss function used for logistic regression. Here 10 hidden neurons were used to have a reasonably large function space and two layers were used to learn the representations. Building on this, the number of layers can also be increased and a comparison can be done on the basis of a chosen metric. From initial analysis, with two layers, the performance on the training set and the test set was fairly similar. Also, stochastic gradient descent was used with Nesterov Accelerated Gradient\cite{noauthor_optimization_nodate} and a momentum value of 0.9. Such a setting was used to accelerate the learning process. In addition, other optimization procedures can be used for the training process and a comparative analysis can be carried out. 100 epochs were used with a batch size of 10 samples.\\
The meta-parameter under consideration is the learning rate and the values under observation are:
\begin{enumerate}
    \item 0.001
    \item 0.01
    \item 0.1
\end{enumerate}
\subsubsection{Statistical Significance Tests}
Statistical Significance Tests are used to analyze any two models in terms of measures of accuracy or error.

The paired t-test can be used to compare the models under observation which is given by the following.
\begin{align}
t = \frac{\Bar{x} - \mu_0}{s / \sqrt{n}}
\end{align}
where $\Bar{x}$ is the sample mean, $\mu_0$ is the population mean, s is the sample standard deviation, and n in the sample size. \\
However, the t-test will not be used for the experiment because due to cross-validation, the assumption that the samples are i.i.d is violated. The reason for this is that the estimated scores are now dependent. The k-fold cross validation will lead to optimistic scores and result in a higher type-1 error\cite{brownlee_statistical_2018}. \\
Now that we understand why we do not use the t-test, we will use the McNemar's test\cite{brownlee_how_2018} for comparing the models under observation. Some reasons for using McNemar's test are given below. It is a distribution-free test and it is suitable for binary classifiers with cross-validation. It must be noted that the McNemar's test does not say which of the two models is better than the other but says whether the two models agree or disagree in the same way or not. An The McNemar's test uses a contingency table to find out the disagreement between the pair of algorithms being compared. A sample contingency table is given in table \ref{tab:contingency_table}
\begin{table}[h!]
    \centering
    \begin{tabular}{||c c c||}
    \hline \hline
         & Classifier 2 correct & Classifier 2 incorrect \\
         \hline
        Classifier 1 Correct & a & b \\
        \hline
        Classifier 2 Incorrect & c & d \\
        \hline
        \hline
    \end{tabular}
    \caption{Contingency table}
    \label{tab:contingency_table}
\end{table}\\
The McNemar's test statistic is given as
\begin{align}
    \chi^2 = \frac{(b-c)^2}{b+c} 
\end{align}
which gives you the value of the statistic, which can be used to retrieve the p-value.
As evident, this test tries to ascertain the measure of disagreement between the two algorithms as to say how do the algorithms disagree on the data. We use the McNemar's test for three pairs of algorithms.
The McNemar's test will be used for evaluating the models in the analysis section.

\subsection{Preliminary Results}
For each of the algorithm, some preliminary results were calculated which are given below. For each of the algorithms, GridSearch\cite{grid_search_sklearn} with k-fold(k=5) cross-validation was used. For each of the algorithm, the ROC curve is provided below which tells us how good the model is in classification. This is important because we do not just want our model to achieve a high accuracy but also the quality of the model should be good in terms of the number of true positives and false positives. \\ In addition to the ROC curve, the precision-recall curve is also provided because it help us to understand the performance of the model in terms of the number of correct, incorrect predictions, and the total predictions. Even though the dataset under consideration has almost balanced classes, the precision-recall curve can tell us about the relevant predictions returned out of the total number of predictions.
\subsubsection{Logistic Regression}
\begin{figure}[h!]
    \centering
    \includegraphics[width=0.6\linewidth]{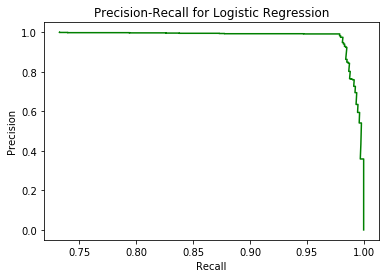}
    \caption{Precision-Recall curve for Logistic Regression}
    \label{fig:pr_lr}
\end{figure}

With k-fold cross-validation, and k=5, 1000 iterations, and a tolerance level of 0.00001, the best parameter that was obtained was the \textbf{inverse regularization}\cite{noauthor_sklearn.linear_model.logisticregression_nodate} of 0.1.
The accuracy that was obtained with L2 regularization was 95.72(with a standard deviation of 1.325\%) without feature scaling (for the mentioned number of iterations and the tolerance level) and an accuracy of 98.39(with a standard deviation of 0.55\%) with feature scaling. This is because the gradient takes longer time to converge because the contours for the features are not symmetrical in nature. Also, the rest of the algorithms will be trained and tested on the scaled dataset. \\
Now that we had a chance to look at the accuracy, let's understand how good our model is in separating the two classes of the problem. To understand this, we will have to look at the ROC curve and the Area under the Receiving Operating Characteristics curve.
\begin{figure}[h!]
    \centering
    \includegraphics[width=0.6\linewidth]{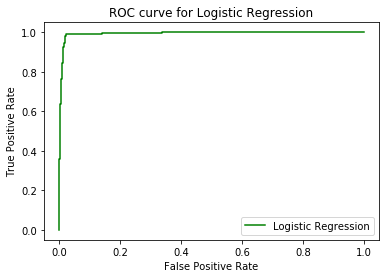}
    \caption{ROC curve for Logistic Regression}
    \label{fig:roc_lr}
\end{figure}
\\
From figure \ref{fig:roc_lr}, it is very clear that the model when very well classify between the two classes. \\
Correspondingly, the area under the receiving operating characteristics curve achieved a score of 99.41.
The precision recall curve for logistic Regression is given in figure \ref{fig:pr_lr}
\newpage
\subsubsection{Decision Tree}
\begin{figure}[h!]
    \centering
    \includegraphics[width=0.6\linewidth]{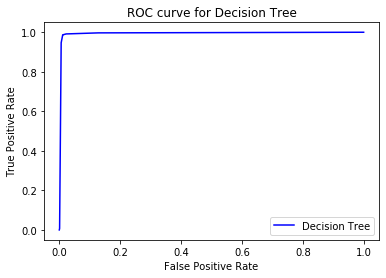}
    \caption{ROC curve for Decision Tree}
    \label{fig:roc_dt}
\end{figure}

The Decision Tree algorithm was also used with GridSearch to determine the best parameters. The meta-parameter under consideration was the max\_depth and the best value for max\_depth was found to be 6 when using k-fold cross-validation(k=5). \\
When using the chosen best parameter max\_depth as 6, an accuracy of 98.53\%(with a standard deviation of +- 0.35\%) is obtained with the feature scaled dataset.
The receiving operating characteristics curve is shown in figure \ref{fig:roc_dt}.
 \\
The area under the curve for the roc curve achieved a score of 99.39\% which is almost similar to that of Logistic Regression.

\begin{figure}[h!]
    \centering
    \includegraphics[width=0.6\linewidth]{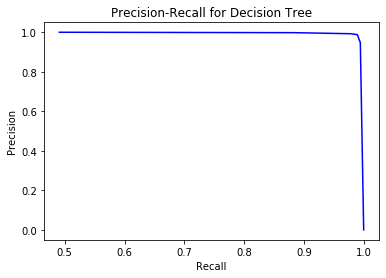}
    \caption{Precision-Recall curve for Decision Tree}
    \label{fig:pr_dt}
\end{figure}

The precision recall curve for decision tree is given in figure \ref{fig:pr_dt}

\newpage

\subsubsection{Neural Network}
\begin{figure}[h!]
    \centering
    \includegraphics[width=0.6\linewidth]{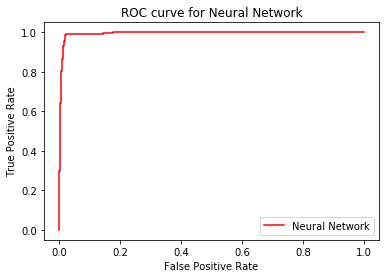}
    \caption{ROC curve for Neural Network}
    \label{fig:roc_nn}
\end{figure}

As mentioned before, the neural network contained two hidden layers with a rectified linear unit activation in the hidden layers. Using GridSearch, the best value of the learning rate was found to be 0.001 with a batch size of 10, and for 100 epochs. The average accuracy was 98.33\% (with a standard deviation of 0.53\%) over 10 runs, for the learning rate of 0.001 with a batch size of 10 for 100 epochs. The neural network used a momentum of 0.9 with Nesterov Accelerated Gradient to decrease the learning time. In addition, the Adam optimizer was also used to train the network which gave a mean accuracy of 98.37\%  +-  0.6\%. Since there is no significant difference in the estimates of accuracy for both the optimizers, we select the Nesterov Accelerated Gradient as the choice of optimizer on the basis of lower standard deviation. Further, more analysis can be carried out in terms of comparing different optimizers for the experiment and their effects on the prediction performance. However, for this experiment such analyses is out of scope for the problem.
The receiving operating characteristics curve for the neural network is shown in fig \ref{fig:roc_nn}.
\\
The area under the roc curve also received a score of 99.49\% which is slightly higher when compared to other models in the experiment.

\begin{figure}[h!]
    \centering
    \includegraphics[width=0.6\linewidth]{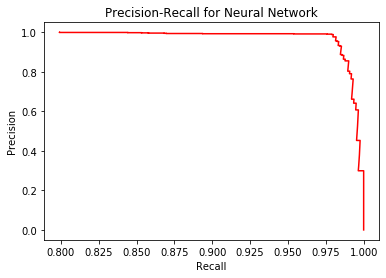}
    \caption{Precision-Recall curve for Neural Network}
    \label{fig:pr_nn}
\end{figure}

The precision recall curve for neural network is given in figure \ref{fig:pr_nn}
\newpage
\section{Analysis}
All the three algorithms will be analyzed in terms of the following:
\begin{enumerate}
    \item Execution time
    \item Statistical Significance Test
    \item Prediction Performance
\end{enumerate}
\subsubsection{Execution time}
Logistic Regression took an average of 5.39 seconds for the training process for the non-scaled dataset and an average of 0.06 seconds to train on the scaled dataset. This shows that the training process is expedited when the dataset is scaled. This behavior is observed because the contours of the features are of different scales and thus the gradient takes uneven steps to reach the minima and thus takes longer in terms of execution time. This result also shows that we must always scale are dataset while training complex models like neural networks because they inherently have longer execution times. \\ \par
Decision Tree provides interesting results compared to other logistic regression because of the selection meta-parameters. Using GridSearch with the scaled dataset, we found that the optimal depth of the decision tree is six. Now, it is obvious that since the depth of tree is fixed, the time required for the training process will be the same, which, on average was 0.08 seconds for a max depth of 6, but the accuracy on the scaled dataset and the non-scaled dataset will be different. In addition, it is worthwhile to report that when using cross-validation with the scaled dataset, the max depth meta-parameter of the tree is found out to be six, but when using cross-validation with the non-scaled dataset, the max depth was found to be 18. For memory considerations and execution time, lower tree depth is beneficial for the training process. \\ \par
For the two hidden layered neural network, the training time with the scaled dataset was about 326.33 seconds on average (with a batch size of 10, momentum of 0.9 and using Nesterov Accelerated Gradient). Increasing the size of the neural network did not result in considerable increase in the performance of the model and therefore the two layer neural network is selected here. On the other hand, training a neural network with a single hidden layer resulted in a lower training time on average but with a cost of a minor drop in accuracy (about 1\%). In addition, using the the Adam
\subsubsection{Statistical Significance Test}
\label{sec:stats_analysis} 
Following the explanation from the experimental design section, we will use the McNemar's test for comparing the algorithms for the binary classification problem in this experiment. \\
It must be noted that the McNemar's test does not say which algorithm is better than the other but says whether the two algorithms under consideration disagree in the same manner or not. This is important because there is no significant difference in the accuracy of the models on the test data for multiple runs.
The null hypothesis for each pair of the three algorithms is given below. The value for $\alpha$ is set to 0.05, which means that 95\% of the time, we reject the null hypothesis. Also, it must be noted that if the sum of disagreements is less than 25 the binomial distribution is used, which is the  default behaviour for this experiment. If the sum is equal to above 25, the distribution changes to a chi-squared distribution.\\
\begin{enumerate}
    \item Logistic Regression = Decision Tree
    \item Decision Tree = Neural Network
    \item Logistic Regression = Neural Network
\end{enumerate}
The alternate hypothesis is that there is significant statistical difference between the pair of algorithms as shown below.
\begin{enumerate}
    \item Logistic Regression $\neq$ Decision Tree
    \item Decision Tree $\neq$ Neural Network
    \item Logistic Regression $\neq$ Neural Network
\end{enumerate}
The algorithms that are different have to be inferred from the statistical significance test. 

\begin{itemize}
    \item Logistic Regression and Decision Tree - The p-value obtained from the test for logistic regression and decision tree is 0.012 on average which is less than the critical value of $\alpha=0.05$. Therefore, we conclude that the null hypothesis is rejected and there is significant difference in the disagreement between the two algorithms. Looking the contingency table, logistic regression gets more predictions incorrect when decision tree gets the predictions correct compared to the other case where the incorrect number of predictions is less for decision tree where logistic regression gets them correct. This says that the decision tree algorithm performs slightly better than logistic regression. However, the differences in the disagreements are significant and therefore the are statistically different.
    \item Decision Tree and Neural Network - The p-value obtained from the McNemar's test is 0.0044 which is lower than the value of $\alpha=0.05$. Therefore, there is significant difference between the a Decision Tree and Neural Network. Looking at the contingency matrix, we observe that 21 times the decision tree algorithm got correct number of predictions when the neural network predicted the observations incorrectly. In addition, 3 times, the decision tree got the observations incorrect when the neural network predicted them correctly. Again, this indicates that the decision tree algorithm performs better for the experiment under consideration.
    \item Logistic Regression and Neural Network - The p-value obtained in this case was 0.015 when considering a neural network trained with a batch size of 10 and 0.5 when trained with a batch size of 100. When training with a batch size of 10, the logistic regression performed slightly better but the difference is not significant because the number of incorrect predictions for the neural network when logistic regression predicted correctly was 7 times and there were no instances where logistic regression predicted incorrectly and the neural network predicted correctly. In addition to this result, when using a batch size of 100, the p-value of 0.5 was obtained which is over the value of $\alpha=0.05$, which says that there is a significant difference. This also agrees with the contingency table of logistic regression and the neural network (trained with a batch size of 100 epochs), which does not show a significant difference between between the number of incorrect predictions for logistic regression and neural network. \par
    In order to save resources, time, and capital, the decision tree algorithm should be considered while classifying kepler objects of interests either as candidate exoplanets or false positives,
\end{itemize}
The results for the McNemar's test are summarized in table \ref{table:1} (with a neural network trained on batch size of 100)
\begin{center}
\begin{table}[h!]
 \begin{tabular}{||c c c||} 
 \hline
 Algorithm 1 & Algorithm 2 & P-Value \\ [0.5ex] 
 \hline\hline
 Logistic Regression & Decision Tree & 0.012 \\ 
 \hline
 Decision Tree & Neural Network &  0.0044 \\
 \hline
 Logistic Regression & Neural Network & 0.5 \\
 \hline
\end{tabular}
\caption{P-values for the algorithm pairs}
\label{table:1}
\end{table}
\end{center}
From the values in table \ref{table:1}, we can infer that the third null hypothesis is true which states that there exists no statistically significant difference between the performance of Logistic Regression and the Neural Network. Again, it is emphasized that the McNemar's test does not state which algorithm is better than the other, but says to what extent do the algorithm disagree with each other. \\ \par
The decision of rejecting or accepting the null hypothesis is given in \ref{table:results}.
\begin{center}
\begin{table}[h!]
 \begin{tabular}{||c c c||} 
 \hline
 Algorithm 1 & Algorithm 2 & Reject Null Hypothesis? \\ [0.5ex] 
 \hline\hline
 Logistic Regression & Decision Tree & Yes \\ 
 \hline
 Decision Tree & Neural Network &  Yes \\
 \hline
 Logistic Regression & Neural Network & No \\
 \hline
\end{tabular}
\caption{Hypothesis decision results}
\label{table:results}
\end{table}
\end{center}
\subsubsection{Prediction performance}
Instead of looking at the actual numbers of false positives and false negatives, a closer look at the precision-recall curve for each algorithm provides some good insights.\par For logistic regression, the precision drops steeply when the recall goes over around 98\%. This means that there is low chance that the predictions of logistic regression will be relevant when the goal is to improve the accuracy of the model. In other words, an increase in the recall results in some of the samples to be incorrectly predicted; for example some samples are predicted to belong in class 0, when the actual prediction is that of class 1. This is not desirable because classifying a kepler objects of interest is important as only when the planet is classified as a candidate planet, only then the space organization should invest time and funds for further investigations. \par
For the decision tree algorithm, the precision-recall curve shows that when the recall reaches almost a 100\%, only then the precision drops steeply. This shows that the observations are predicted almost 100\% correctly and the samples that are predicted correctly are relevant. In other words, the samples that are predicted to be in class 1 have an extremely low chance of being in class 0 and vice versa. This result is better than that of logistic regression. When comparing logistic regression to the decision tree algorithm, the choice of algorithm for the dataset would be decision tree. \par
For the neural network, when trained with a batch size of 10, the accuracy is the comparable to other models. However, from the precision-recall curve, we can observe that the precision drops sharply when the recall goes just over 97.5\%. The amount of relevant predictions is still small but the difference is significant when compared to the decision tree algorithm. Again, the observation from the precision-recall curve for the neural network agrees with the fact that decision tree will be a good algorithm to choose when considering the scaled dataset and meta-parameters selected in this experiment from cross-validation. \par It is surprising to see that the neural network performs slightly poorly than decision tree and logistic regression in terms of precision and recall even when the accuracy is on the higher side. Therefore, considering only accuracy as the metric for algorithm selection will be insufficient when there is no significant difference over multiple runs. From the results above, it can be seen that decision tree outperforms logistic regression and the two-layered neural network in terms of predicting accurately and relevantly.
\section{Conclusion}
The project conducted an experiment on a dataset that contained information about kepler objects of interests. The dataset contained features observed from the kepler satellite. Using this dataset, basic exploratory data analysis helped to visualize the data and get a high level understanding of the data. The selection of the machine learning models were carried out depending on existing works on binary classification with the hope that the selected models will perform comparably to other experiments with similar conditions. The models were run with cross-validation to help select the best meta-parameters for each model. The accuracy for each algorithm is evaluated against each other as basic comparison technique. In addition, the receiving operating characteristic and the precision-recall curves were plotted to understand the diagnostic ability of the machine learning model and the number of relevant results returned from the model during the prediction, respectively. The algorithms were also compared using the McNemar's test which led to rejecting the null hypothesis on two occasions and accepting the null hypothesis on one occasion. Ultimately from the statistical significance tests, accuracy, and the execution time, it can be concluded that the decision tree algorithm with the selected meta-parameter(max\_depth$=$6) was the model that performed optimally.
\section{Future Work}
As an extension to the experiment, it would be beneficial for this problem to study the effects of different effects of using various optimization algorithms for the classification process. Especially for the neural network, it will be interesting to study the effects of choosing various optimization algorithms with respect to increasing the number of layers and its effect on the performance. Other parameters can also be tested, which will be critical to classification of astronomical objects. For example, feature engineering techniques can be used to create new features which can help to learn other attributes from the data and thus improving the models performance. Engaging in such studies will help to build and use sophisticated models in analysing various astronomical objects. 
\printbibliography
\end{document}